# High gain backward lasing in air

Arthur Dogariu,[1,*] James Michael,[1] Marlan O. Scully,[1,2] and Richard B. Miles[1]

[1]*Mechanical and Aerospace Engineering Department, Princeton University, Princeton, NJ 08544,*
[2]*Institute of Quantum Science and Engineering, Texas A&M University, College Station, TX 77843*
[*]*adogariu@princeton.edu*

The need for molecular standoff detection has motivated the development of a remotely pumped, high gain air laser that produces lasing in the backward direction and can sample the air as the beam returns. High gain is achieved in the near infrared following pumping with a focused ultraviolet laser. The pumping mechanism is simultaneous resonant two-photon dissociation of molecular oxygen and resonant two-photon pumping of the atomic oxygen fragments. The high gain from the millimeter length focal zone leads to equally strong lasing in the forward and backward directions. Further backward amplification is achieved using prior laser spark dissociation.

## 1. Introduction

Optical techniques for the remote detection of atoms and molecules rely on the use of lasers to selectively identify and quantify species of interest. To enable single-sided detection, collection of light must be accomplished in the backward direction. Collection of incoherent light emission from molecules of interest is limited by the non-directional nature of spontaneous emission. More sensitive detection techniques, aided by the coherent nature and well-defined direction of emission, are restricted in the direction of emission by the phase-matching relationship. For commonly employed nonlinear techniques such as coherent anti-Stokes Raman spectroscopy (CARS) [1] and stimulated Raman scattering (SRS) [2], phase-matching results in a coherent beam propagating in the direction of the pumping laser, *away* from the source.

These limitations have motivated the exploration of backward air lasing and stimulated gain concepts, which can produce coherent scattering that returns to the pump laser location [3]. To date, the only approach that has shown promise is based on the electron recombination of ionized molecular nitrogen from a femtosecond produced filament [4,5]. This scheme leads to gain at 337 nm, the same wavelength as the discharge-pumped molecular nitrogen laser. Amplified spontaneous emission gain on the order of 0.3 cm$^{-1}$ has been observed [5].

Here, we demonstrate the generation of high gain lasing in air using a 100 picosecond remote pump laser, which simultaneously drives a two-photon dissociation of molecular oxygen and a two-photon excitation of one of the resulting oxygen atom fragments. Both processes are resonantly enhanced at the 226 nm wavelength of the pump laser. The excitation is followed by lasing from the excited atomic oxygen as shown in Figure 1A. The pump laser is focused such that there is no laser induced breakdown of the air, and excitation followed by stimulated emission is achieved throughout the 1 mm long focal region. The result is the formation of well collimated backward and forward propagating laser beams at 845 nm with parameters corresponding to the ultraviolet pump beam focusing.

## 2. Atomic oxygen generation

Two-photon laser induced fluorescence from atomic oxygen has been developed for the quantitative diagnostics of combusting gases where atomic oxygen is an important radical species [6,10]. The two-photon excitation transition is from the $2p^3P$ ground state to the $3p^3P$ excited state using 226 nm laser radiation. That excitation is followed by spontaneous relaxation from the $3p^3P$ state to the $3s^3S$ state, producing fluorescence emission at 845 nm (see Fig. 1B). The use of the two-photon excitation to produce stimulated emission at 845 nm in atomic oxygen has been observed in flames at sub-atmospheric pressures [11].

The same two-photon transition can be used as the initial step in a 2+1 Resonant Enhanced Multiphoton Ionization, or REMPI [12]. This process can be remotely monitored by microwave scattering from the free electrons (Radar REMPI [13]). Fig. 1C shows the Radar REMPI excitation spectra from a flame containing oxygen atoms (squares) and from ambient air (circles) using a 100 picosecond (ps) laser tuned in wavelength through the two-photon oxygen atom transition at 226 nm. The 100 ps laser pulse is short enough to suppress avalanche breakdown [14], so multiphoton ionization of atomic oxygen dominates and the measured Radar REMPI signal reflects the density of atomic oxygen in the focal region of the laser. Note that in room air, even in the absence of the flame, the Radar REMPI signal from atomic oxygen is present, indicating that the laser has caused the dissociation of molecular oxygen. Similar molecular oxygen dissociation and atomic oxygen excitation has previously been observed with nanosecond pulses using two-photon laser induced fluorescence [8,11].

molecular oxygen photodissociation mechanisms and atomic oxygen fragment velocities [15]. Their results indicate that three dissociation pathways are present at 226 nm: (i) a quite weak single photon dissociative path via the Herzberg continuum forming $O(^3P_2) + O(^3P_j)$ with 0.38 eV excess energy, (ii) a dominant two-photon dissociative path via a resonance at 225.66 nm with the predissociative $3d\delta\ ^3\Pi_{0,1}(v=2)$ Rydberg state leading to $O(^3P_2) + O(^1D_2)$ with 3.91 eV excess energy, and (iii) a weak two-photon resonant pathway through the same Rydberg state leading to $O(^3P_2) + O(^3P_j)$ with 5.88 eV excess energy . The dominant resonant two-photon pathway produces $O(^3P_2)$ atoms with velocities on the order of 4800 m/s.

Our previous observation that the 100 ps laser pulse at 226 nm can simultaneously dissociate the molecular oxygen to ground state atoms and excite the fast moving oxygen atoms [16] led us to the realization that lasing might be possible in room temperature atmospheric pressure air. By increasing the focal depth of the pump laser and reducing the peak intensity, the photodissociation process takes place without measurable ionization or spark formation, so the pump laser is not scattered by plasma effects and the gain path length of the atomic oxygen transition reflects the focal depth of the pump laser.

## 3. Experimental Results and Discussion

The experimental set up is arranged to measure the gain and lasing properties in the backward direction, toward the pump source. The backward lasing is separated from the oppositely propagating pump by a dichroic filter and monitored either by a detector, a fast-gated CCD camera, or a spectrometer. The pump laser is focused into air with a 30 cm focal length lens, leading to a two-photon excitation volume of approximately 10 microns diameter and 1 mm length.

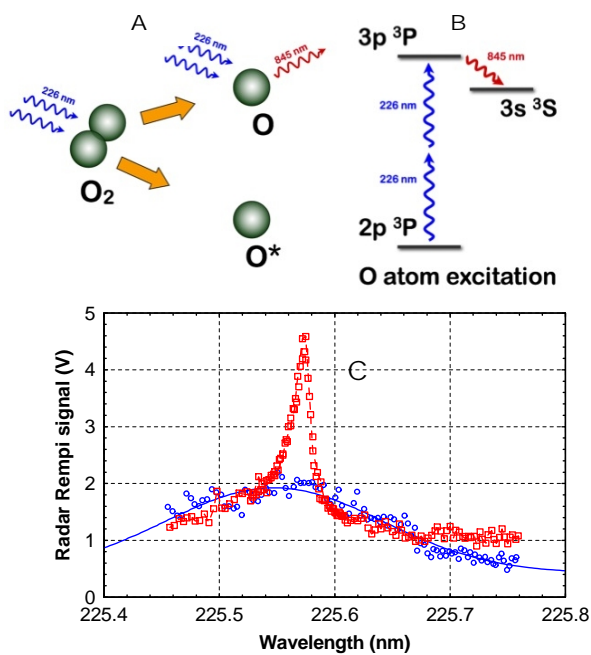

Fig. 1. (A) Two-photon dissociation of the oxygen molecule and subsequent two-photon resonant excitation of the ground $2p^3P$ state oxygen atom fragment results in emission at 845 nm. (B) The two-photon excitation of the atom uses two 226 nm photons to excite from $2p^3P$ to $3p^3P$ which either fluoresces or lases to the $3s^3S$ level, emitting a photon at 845 nm. (C) Atomic oxygen 2+1 Radar REMPI signals in a flame (squares) and ambient air (circles) demonstrate the atomic oxygen resonance linewidth with preexisting oxygen atoms in the flame and with photo dissociation produced oxygen atoms in air.

The Radar REMPI spectra show that the bandwidth of the atomic oxygen resonance seen in air is approximately ten times wider than the atomic oxygen bandwidth seen in the flame, indicating that the signal is coming from high velocity oxygen atom fragments. The high velocity of the fragments suggests that they are formed by two-photon rather than single photon dissociation. This observation is consistent with the measurements of Buijsse, et al. who utilized a 2+1 REMPI time of flight ion imaging scheme to measure

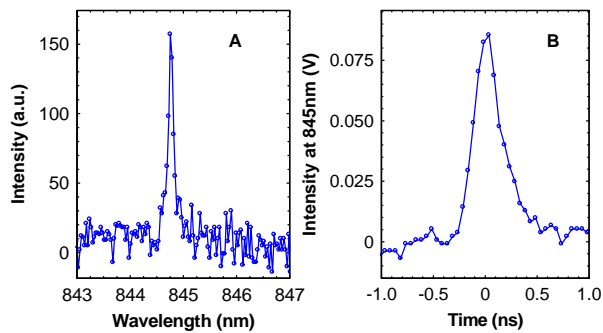

Fig. 2. Backscattered atomic oxygen emission (A) measured spectrum and (B) pulse in time.

Fig. 2 shows the measured spectrum and temporal pulse shape of the backscattered emission from atomic oxygen at 845nm. The atomic oxygen spectrum shown in Fig. 2A is much narrower than the 0.1 nm resolution limit of the spectrometer. The 3GHz oscilloscope used to measure the backscattered pulse in Fig. 2B indicates an upper limit of 300ps for the atomic oxygen emission. The fact that the backwards emission pulsewidth is more than two orders of magnitude shorter than the reported atomic oxygen fluorescence lifetime [11] of 36 ns is one indication of stimulated emission.

In order to further establish the presence of lasing, the signal detected along the axis of the volume is compared to the signal detected observing from the side. Since both approaches observe the same excited volume and collect the same solid angle, if the signal is incoherent fluorescence, the two measurements should give similar results. On the other hand, if the signal is due to stimulated emission, then there is exponential gain and the path length is important. In that case, the signal seen emanating from the volume along the axis will be significantly stronger that that seem from the side. Using a photomultiplier tube we were able to measure a ratio of more than 8000 between the emission collected within the same 0.78 steradian (sr) solid angle along the axis in the backscattered direction and that collected from the side [see Fig. 3**A**,**B**], indicating that the backscattered emission is strongly gain dominated.

Furthermore, if the emission were primarily incoherent, the scaling of the detected emission with the input UV pump laser energy would be independent of the direction of observation. Fig. 3 shows that the scaling is dramatically different. The axial signal (A) scales as the 4.2 power of the pump laser and has a threshold value, whereas the side view signal (B) scales as the 2.3 power of the pump laser and has no threshold. In previous work employing nanosecond laser pumping, the two-photon fluorescence scaling in air has been observed to be ~4 (*11*) rather than 2.3, however in the experiments reported here the intensity of the pump laser is two orders of magnitude higher. Experiments with the same focusing in a stoichiometric methane/air flame where atomic oxygen is naturally present showed strong saturation of the two-photon pumping of the atomic oxygen. Thus, we conclude that for the 100 ps pumping laser in air the two-photon excitation of atomic oxygen is saturated. The 2.3 power scaling with the pump energy indicates that the formation of atomic oxygen from two-photon dissociation of molecular oxygen does not exhibit saturation. Furthermore, in Fig. 3C a direct comparison between the backwards emission and side emission (while varying the pump power) shows a nonlinear dependence with the power of 2. Since the voltage measurements of the two photomultipliers can be directly related to the respective signal amplitudes, from Fig. 3C we can also estimate the optical gain seen by the axial beam due to the stimulated emission process: the backscattered beam (which can be all collected in a 0.006 sr solid angle) is about 500 times stronger than the overall incoherent emission (obtained by extending the 0.78 sr side collection efficiency over the whole $4\pi$ solid angle).

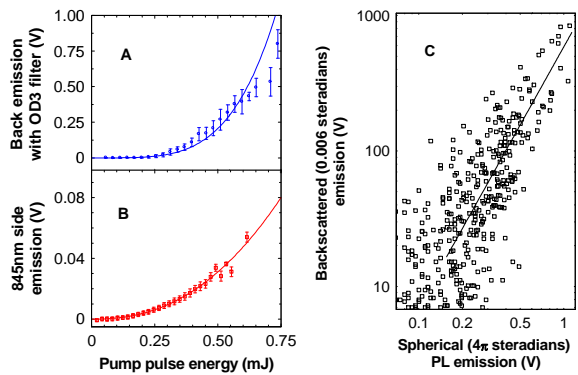

Fig. 3. The backward (A) and side (B) 845nm emission in air as function of the 226nm pump laser energy shows the scaling difference between backward lasing (4th power) and incoherent emission (2nd power), respectively. The curves represent a 4.2 power fit offset by a 0.2 mJ threshold for the backward emission(A) and a 2.3 power fit through the origin for the side emission (B). The backwards coherent emission versus the total non-directional incoherent emission (C) shows the stimulated emission gain. The line in (C) has a slope of 2 indicating a quadratic scaling ratio of backward coherent to total incoherent emission.

This ratio of 500 between the energy of the backscattered laser and the total fluorescence over the 1 mm path length corresponds to an optical gain coefficient of 62 cm$^{-1}$. An overall signal gain in the backscattered direction can be estimated by using the ratio of the laser energy to the energy of the fluorescence collected into the same solid angle as the laser beam (6 x 10$^{-3}$ sr). The solid angle ratio coupled with the optical gain of 500 energy increase of the axial beam when compared to the total fluorescence leads to an increase on the order of 10$^6$ along the 1 mm axial path length of the excitation volume. Similar gain is measured in the forward direction.

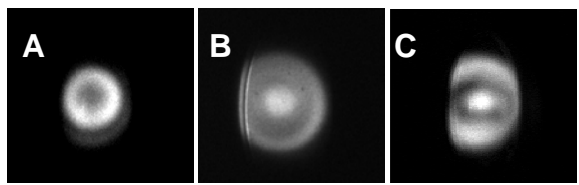

Fig. 4. Backscattered oxygen laser beam at 845nm measured 30cm away from the 226nm laser focused in (A) air, (B) in a methane-air flame, and (C) in air with a 532nm pre-pulse generated spark on the far side of the gain region.

As a final confirmation of lasing, Fig. 4A shows the image of the backward propagating lasing spot seen by a CCD camera. Although it has a donut shape, the beam is well localized and the divergence of 40 mradians is consistent with diffraction limited lasing from the cross sectional size of the pump volume. The measured energy of the atomic oxygen laser pulse obtained in air is ~10 nJ.

The observed donut-shaped mode is related to the dissociation of molecular oxygen. The presence of pre-existing atomic oxygen generated by other means leads to a much stronger signal and produces a mode with a central maximum. For example, Fig. 4B shows the image of backward lasing from naturally occurring atomic oxygen in an atmospheric pressure, stoichiometric methane/air flame, while Fig. 4C shows the atomic oxygen lasing from air when a spark is induced a few microseconds earlier by a 532nm pulse from a frequency doubled Nd:YAG laser. In both of these cases the lasing is stronger and better collimated due to the higher gain associated with the higher volume fraction and lower velocity of the atomic oxygen. Using the Nd:YAG pre-pulse to create atomic oxygen enhances the backscattered emission by more than two orders of magnitude, and can be successfully used to improve the coherent emission from the remotely induced air laser. For that enhancement to be effective for stand-off applications, the spark must be located just on the far side of the lasing volume in order to minimize the distortions of the backward propagating laser beam induced by the high temperature spark region.

## 4. Conclusions

In summary, we have achieved high gain lasing in air through the two-photon dissociation of molecular oxygen and the subsequent two-photon pumping of the atomic oxygen fragments. A well collimated laser beam propagating in the forward and backward directions is formed indicating an optical gain well in excess of 60 cm$^{-1}$. The divergence of the generated laser beams is less than 40 mradians, consistent with the scale of the pumping volume cross section. The optical gain and directional emission allows for six orders of magnitude enhancement for the backscattered emission when compared with the fluorescence emission collected in the same solid angle as the stimulated emission. This achievement opens new opportunities for the remote detection of atmospheric contaminants and trace species through stimulated and nonlinear backward scattering processes.

## 5. Acknowledgements.

We gratefully acknowledge the financial support from the Office of Naval Research under the Science Addressing Asymmetric Explosive Threats (SAAET) Program.